\documentclass[pra,aps,amsmath,amssymb,twocolumn,showpacs]{revtex4}
\usepackage{graphicx}

\newcommand{\aver}[1]{\left< #1 \right>}
\newcommand{\ket}[1]{\left| #1 \right>}
\newcommand{\bra}[1]{{\left< #1 \right|}}
\DeclareMathOperator{\Tr}{Tr}

\begin{document}
\title{On the quantum (in)stability in cavity QED}
\author{S.V.~Prants and M.Yu.~Uleysky}
\affiliation{Laboratory of Nonlinear Dynamical Systems,\\
V.I. Il'ichev Pacific Oceanological Institute of the Russian Academy of
Sciences,\\
690041 Vladivostok, Russia}
\date{\today}
\begin{abstract}
The stability and instability of quantum motion is studied in the context
of cavity quantum electrodynamics (QED). It is shown that the Jaynes-Cummings
dynamics can be unstable in the regime of chaotic walking of an atom in the
quantized field of a standing wave in the absence of any other interaction 
with environment. This quantum instability manifests itself in strong variations
of quantum purity and entropy and in exponential sensitivity of fidelity of quantum
states to small variations in the atom-field detuning. It is quantified in terms
of the respective classical maximal Lyapunov exponent that can be estimated in
appropriate in-out experiments.
\end{abstract}
\maketitle

The problem of stability of quantum dynamics has attracted a great interest by its
own right and in relation to the field of quantum information and computation.
Classical instability is usually defined as an exponential separation of two
nearly trajectories in time with an asymptotic rate given by the maximal Lyapunov
exponent $\lambda$. Perfectly isolated quantum systems are unitary and cannot be unstable
in this sense even if their classical limits are chaotic \cite{C91}. It is well
known that quantum coherence is destroyed due to interaction with an environment
\cite{Z91,G96} which is usually modeled by a heat bath with infinitely many
degrees of freedom. Environment-induced decoherence causes quantum-entropy increase
which is dominated by the classical Lyapunov exponents \cite{KZ03}. In an
alternative approach \cite{Peres,Caves} the quantum instability is proposed to
be measured by the decay of the fidelity or the overlap 
\begin{equation}
f(\tau)=\left|\left<\Psi_1(\tau)\right.\left|\Psi_2(\tau)\right>\right|^2
\label{Fidelity}
\end{equation}
of two wave functions $\Psi_1$ and $\Psi_2$, identical at $\tau=0$, that evolve under
slightly different Hamiltonians.

In a number of numerical studies \cite{JP01,JSB01,BC02,PSZ03} for a variety of
classically chaotic models it has been established that the overlap decay may be 
algebraic, Gaussian, and exponential. The strength of perturbations in Hamiltonians
and other factors determine which of these regimes prevails.

In this letter we show that instability of quantum dynamics and its exponential
sensitivity to initial conditions and small variations in parameters may occur in
a paradigmic cavity-QED system with a single environmental degree of freedom.
To specify the problem we consider the standard model in cavity QED
Jaynes-Cummings Hamiltonian \cite{JC}
\begin{multline}
\hat H=\frac{\hat p^2}{2m_a}+\frac{1}{2}\hbar\omega_a\hat\sigma_z+\hbar\omega_f\hat a^\dag\hat a-\\
-\hbar\Omega_0\left(\hat a^\dag\hat\sigma_-+\hat a\hat\sigma_+\right)\cos{k_f\hat x},
\label{Janes-Cum}
\end{multline}
which describes the interaction between a two-level atom with lower, $\ket{1}$, and
upper, $\ket{2}$, states, the transition frequency $\omega_a$, and the Pauli operators $\hat\sigma_{\pm,z}$
and a quantized electromagnetic-field mode with creation, $\hat a^\dag$, and annihilation,
$\hat a$ operators forming a standing wave with the frequency
$\omega_f$ and the wave vector $k_f$ in an ideal cavity. 
The atom and field become dynamically entangled by
their interaction with the state of the combined system after the interaction
time $t$
\begin{equation}
\ket{\Psi(t)}=\sum_{n=0}^\infty a_n(t)\ket{2,n}+b_n(t)\ket{1,n}
\label{Psi}
\end{equation}
to be expanded over the Fock field states $\ket{n}$, $n=0,1,\dots$. Here 
$a_n(t)\equiv\alpha_n(t)+i\beta_n(t)$ and $b_n(t)\equiv\rho_n(t)+i\eta_n(t)$ are the
complex-valued probability amplitudes to find the field in the state $\ket{n}$ and the
atom in the states $\ket{2}$ and $\ket{1}$, respectively. In the process of emitting and
absorbing photons, atoms not only change their internal electronic states
but their external translational states change as well due to the photon
recoil effect. If atoms are not too cold and their average momenta are large
as compared to the photon momentum $\hbar k_f$, one can describe the translational degree
of freedom classically. The whole dynamics is now governed by the Hamilton-Schr\"odinger
equations \cite{PU03} that have the following normalized form in the frame rotating
with the frequency $\omega_f(n+1/2)$:
\begin{equation}
\begin{aligned}
\dot x&=\kappa p,\\
\dot p&=-2\sin x\sum_{n=0}^\infty\sqrt{n+1}\left(\alpha_n\rho_{n+1}+\beta_n\eta_{n+1}\right),\\
\dot\alpha_n&=-\frac{\delta}{2}\beta_n-\sqrt{n+1}\eta_{n+1}\cos x,\\
\dot\beta_n&=\frac{\delta}{2}\alpha_n+\sqrt{n+1}\rho_{n+1}\cos x,\\
\dot\rho_{n+1}&=\frac{\delta}{2}\eta_{n+1}-\sqrt{n+1}\beta_n\cos x,\\
\dot\eta_{n+1}&=-\frac{\delta}{2}\rho_{n+1}+\sqrt{n+1}\alpha_n\cos x,
\end{aligned}
\label{mainsys}
\end{equation}
where $x=k_f\aver{\hat x}$ and $p=\aver{\hat p}/\hbar k_f$ are the atomic center-of-mass position and momentum, respectively.
Dot denotes differentiation with respect to dimensionless time $\tau=\Omega_0 t$, where
$\Omega_0$ is the amplitude coupling constant. The normalized recoil frequency,
$\kappa=\hbar k_f^2/m_a\Omega_0\ll 1$,
and the atom-field detuning, $\delta=(\omega_f-\omega_a)/\Omega_0$, are the control parameters.

Inspite of existence of an infinite number of the integrals of motion
\begin{equation}
R_n=\alpha_n^2+\beta_n^2+\rho_{n+1}^2+\eta_{n+1}^2=\text{const},\quad \sum_{n=0}^\infty R_n\leqslant 1
\label{Rn}
\end{equation}
and conservation of the total energy
\begin{multline}
W=\frac{\kappa p^2}{2}-\frac{\delta}{2}\sum_{n=0}^\infty\left(\alpha_n^2+\beta_n^2-\rho_{n+1}^2-\eta_{n+1}^2\right)-\\
-2\cos x\sum_{n=0}^\infty\sqrt{n+1}\left(\alpha_n\rho_{n+1}+\beta_n\eta_{n+1}\right),
\label{Energy}
\end{multline}
the Hamilton-Schr\"odinger system (\ref{mainsys}) is, in general, non-integrable. The
type of the center-of-mass motion depends strongly on the values of the detuning
$\delta$. In the limit of zero detuning and with initially excited or deexcited
atoms, the optical potential disappears, and atoms move with a constant
velocity $\dot x=\kappa p_0$. The quantum evolution is periodic with the period $\pi/\kappa p_0$, and exact
solutions for purity, von Neumann entropy, fidelity $f(\tau)$, and other quantum
characteristics can be found in the explicit forms. For example, the atomic
population inversion at $\delta=0$ is the following:
\begin{equation}
\begin{gathered}
z(\tau)=\sum_{n=0}^\infty z_n=\sum_{n=0}^\infty z_n(0)\cos{\left(\frac{2\sqrt{n+1}}{\kappa p_0}
\sin{\kappa p_0\tau}\right)},\\
z_n=\alpha_n^2+\beta_n^2-\rho_{n+1}^2-\eta_{n+1}^2.
\end{gathered}
\label{zfromtau}
\end{equation}
With the detuning being large, $|\delta|\gg 0$, the optical potential is shallow, atom
moves with almost a constant velocity, $\simeq\kappa p_0$, slightly modulated by the standing
wave, and its inversion oscillates with a small depth (excepting for the case
of the so-called Doppler-Rabi resonance with maximal Rabi oscillations that
occur at the condition $|\delta|=\kappa p_0$ \cite{UKP03}). If the atomic kinetic energy, $\kappa p^2/2$,
is not enough to overcome barriers of the optical potential, the atomic center
of mass oscillates in one of the potential wells.

\begin{figure}[!htb]
\includegraphics[width=0.45\textwidth]{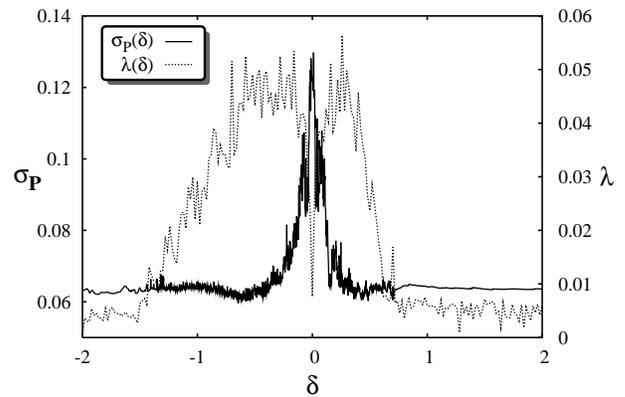}
\caption{Quantum-classical correlation between the dependencies of the
variance of quantum purity, $\sigma_P$ and the maximal Lyapunov exponent $\lambda$ (in units of $\Omega_0$)
on the atom-field detuning $\delta$ (in units of $\Omega_0$).}
\label{Fig1}
\end{figure}
Numerical simulation shows that there exist conditions, defined mainly by the values of the
detuning, when atoms move chaotically in a cavity. This type of motion may be
called a chaotic or random walking, and it is quantified by positive values
of the maximal Lyapunov exponent $\lambda$. In Fig.~\ref{Fig1} we show by the dotted line the 
dependence $\lambda(\delta)$ computed with  Eqs.~(\ref{mainsys}) and the 
following initial conditions: $x_0=0$,
$p_0=25$, the atom is prepared in the state $\ket{2}$ and the field is initially in a coherent
state with the average number of photons $\bar n=10$. The normalized recoil frequency is
chosen to be $\kappa=0.001$, a reasonable value with usual atoms in a high-quality optical
microcavity in the strong-coupling limit. Stability of the computation
with respect to truncating the set (\ref{mainsys}) was checked. In most the cases
$n=100$ was taken. 

The entanglement between the internal atomic and field degrees of freedom can
be characterized by the quantity known as purity 
\begin{equation}
P(\tau)=\Tr_a\rho_a^2(\tau),
\label{Puritydef}
\end{equation}
where $\rho_a(\tau)$ is the reduced atomic density matrix
\begin{equation}
\rho_a(\tau)=\sum_{n=0}^\infty\bra{n}\rho(\tau)\ket{n}
\label{Matrix}
\end{equation}
with the total density matrix to be $\rho(\tau)=\ket{\Psi(\tau)}\bra{\Psi(\tau)}$. Purity is maximal if an atom is in
one of its energetic states $\ket{1}$ or $\ket{2}$; i.~e. $P_\text{max}=\Tr_a\rho_a^2=\Tr_a\rho_a=1$.
Purity is minimal if $\rho_a=I/2$, i.~e. $P_\text{min}=1/2$, where $I$ is the identity matrix. In terms
of the probability amplitudes, it is given by
\begin{multline}
P=\left(\sum_{n=0}^\infty\left(\alpha_n^2+\beta_n^2\right)\right)^2+\left(\sum_{n=0}^\infty\left(\rho_n^2+\eta_n^2\right)\right)^2+\\
+2\left(\sum_{n=0}^\infty\left(\alpha_n\rho_n+\beta_n\eta_n\right)\right)^2-
2\left(\sum_{n=0}^\infty\left(\alpha_n\eta_n+\beta_n\rho_n\right)\right)^2.
\label{Purity}
\end{multline}
The root mean square variance of purity, $\sigma_P=\sqrt{\aver{P^2}-\aver{P}^2}$ has been computed in the range of the
detuning $|\delta|\leqslant 2$ at the same conditions as it was done in computing the maximal
Lyapunov exponent $\lambda$. Irregular oscillations of $\sigma_P$ occurs on the same interval
of $|\delta|\lesssim 1$, where $\lambda>0$ (see Fig.~\ref{Fig1}). Computing the von Neumann entropy,
$S=-\Tr_a(\rho_a\ln\rho_a)$, we have
found the same correlations of its variance with $\lambda$.

\begin{figure}[!htb]
\includegraphics[width=0.45\textwidth]{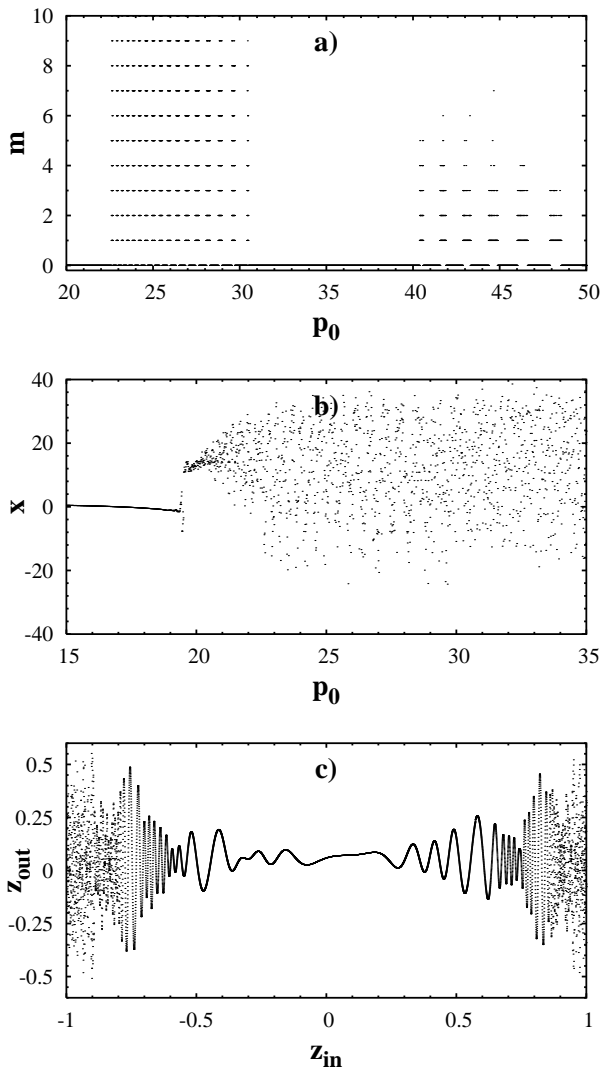}\\
\caption{(a) Fractal set of the initial momenta $p_0$ (in units of $\hbar k_f$) of atoms that
leave a one-wave length cavity after $m$ turns. (b) Sensitive dependence of
the atomic position $x$ (in units $k_f^{-1}$) on the initial momentum $p_0$.
(c) Sensitive dependence of the output values of the atomic population
inversion $z_\text{out}$ on its initial values $z_\text{in}$. 
Control parameters $\delta=0.4$ and $\kappa=0.001$.}
\label{Fig2}
\end{figure}
The chaotic centre-of-mass walking has fractal properties.
Placing atoms at the point $x=0$ with the same initial conditions and parameters but
with different values of initial momenta $p_0$ , we compute the time $T(p_0)$, the atom
with a given value $p_0$ needs to reach one of the nodes of the standing wave at
$x=-\pi/2$ and $x=3\pi/2$, and the number of times, $m$, when it changes its direction of motion. The
scattering function $T(p_0)$ is found to have a self-similar structure with singularities
on a Cantor-like set of initial values of momenta $p_0$. In Fig.~\ref{Fig2}a we demonstrate
the mechanism of generating this set at $\delta=0.4$. There are two sets of atomic 
trajectories with $T\to\infty$, the countable one consisting of separatrix-like trajectories,
corresponding to the ends of the intervals in Fig.~\ref{Fig2}a, and the uncountable
one consisting of trajectories with $m=\infty$. The chaotic motion can be, in principle,
verified in experiments on 1 D-scattering of atoms at the standing wave. Fig.~\ref{Fig2}b
demonstrates sensitive dependence of the atomic positions on $p_0$ 
at a fixed time moment. 
A smooth segment of this function in the range $p_0\lesssim 20$ should be attributed to atomic
oscillations in the first well of the optical potential since these values of $p_0$
are not enough to overcome the respective potential barrier. When $p_0$ exceeds
a critical values, atoms leave the well, and it is practically impossible to
predict even the sign of the atomic position. The so-called predictability 
horizon can be estimated as follows:
$\tau_p\simeq \lambda^{-1}\ln{(\Delta x/\Delta x_0)}$, where $\Delta x$ is the confidence interval and
$\Delta x_0$ the initial inaccuracy in
preparing initial atomic positions. In order to demonstrate the quantum-classical
correspondence in the chaotic regime we compute the dependence of the output
values of the atomic population inversion $z_\text{out}$ at a fixed moment on
its initial values $z_\text{in}$ with the other conditions and parameters to be the same.
Fig.~\ref{Fig2}c illustrates that this function is regular in the range $|z_\text{in}|\lesssim 0.5$, where the
atomic center-of-mass motion is regular, and is irregular if $0.5\lesssim |z_\text{in}|\leqslant 1$ where atoms
move chaotically.

To quantify instability of quantum evolution in cavity QED we compute the
decay of the fidelity $f(\tau)$ which is the overlap (\ref{Fidelity}) of two states $\ket{\Psi_1(\tau)}$ and
$\ket{\Psi_2(\tau)}$,
identical at $\tau=0$, that evolve under two Hamiltonians (\ref{Janes-Cum}) with the slightly
different detunings $\Delta\delta=\delta_1-\delta$.

\begin{figure}[!htb]
\includegraphics[width=0.45\textwidth]{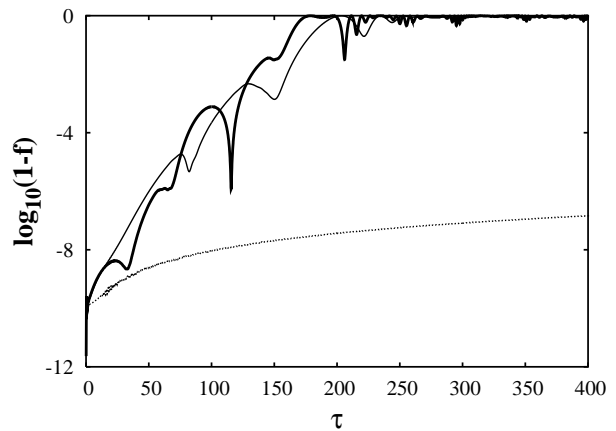}
\caption{Decay of the fidelity of quantum motion $1-f(\tau)$ (logarithmic scale) in the
chaotic (thick and thin lines) and regular (dotted line) regimes. Time $\tau$ is in units of $\Omega_0^{-1}$.}
\label{Fig3}
\end{figure}
We have found previously (see Fig.~\ref{Fig2}) that with the initial momentum $p_0=25$ and $\delta=0.4$ the
type of atomic motion depends strongly on the initial atomic inversion $z(0)$.
If an atom is prepared initially in one of its energetic states, i.~e. $z(0)=\pm 1$, its
classical and quantum dynamics are unstable, whereas they are stable with $z(0)=0$
under the same other conditions. In Fig.~\ref{Fig3} we show for convenience the decay
of the quantity $\log_{10}(1-f)$ in the regimes of chaotic walking (thick and thin lines,
$z(0)=\pm 1$) and regular motion (dotted line, $z(0)=0$) with $\Delta\delta=10^{-4}$. In the chaotic regime the fidelity
decays exponentially with the rate $\lambda\simeq 0.04$ to be equal to the maximal Lyapunov exponent
computed with the set (\ref{mainsys}). This result does not depend on the values of
differences in the control parameters $\Delta\delta$. The fidelity practically does not 
decay in the regular regime at $z(0)=0$, and the respective maximal Lyapunov exponent
was computed to be zero.

We emphasize that sensitive dependence of quantum motion both on initial states
and parameters may arise with an atom in a quantized cavity field. Single
chaotic degree of freedom, arising naturally when we take into account photon
recoils, provides quantum instability and irreversibility. We do not need
an infinite bath or any kind of noise for that. The quantum instability
has been shown to be quantified by the respective maximal Lyapunov exponent
providing a quantum-classical correspondence.
  
\section{Acknowledgments}
The work was supported by the Program ``Mathematical methods in nonlinear dynamics''
of the Russian Academy of Sciences, and by the Far Eastern Division
of the Russian Academy of Sciences.

\end{document}